\documentclass[
    reprint,
    superscriptaddress,
    nofootinbib,
    amsmath,amssymb,
    aps,
]{revtex4-2}

\usepackage{graphicx}
\usepackage{dcolumn}
\usepackage{wasysym}
\usepackage{bm}
\usepackage{euscript}
\usepackage{booktabs}
\usepackage{lineno}
\usepackage{orcidlink}

\defcitealias{PB24}{PB24}
\newcommand\orcid[1]{\href{https://orcid.org/#1}{\includegraphics[height=9pt]{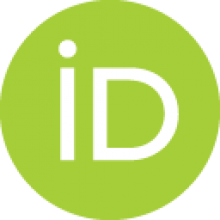}}}

\begin{document}

\preprint{APS/123-QED}

\title{Constraints on the polarization angle oscillations of the Crab Nebula with the\\Simons Array and its applications to the search for axion-like particles}

\author{Tylor~Adkins~\orcidlink{0000-0002-3850-9553}}
\email[Corresponding author: \href{mailto: tadkins@berkeley.edu}{tadkins@berkeley.edu}]{}
\affiliation{Department of Physics, University of California, Berkeley, CA 94720, USA}

\author{Shahed~Shayan~Arani~\orcidlink{0009-0005-4168-2858}}
\affiliation{Department of Physics, University of California, San Diego, La Jolla CA 92093, USA}
\affiliation{Center for Dark Cosmology and Gravitation, University of California, San Diego, La Jolla, CA 92093-0319, USA}

\author{Kam~Arnold~\orcidlink{0000-0002-3407-5305}}
\affiliation{Department of Astronomy \& Astrophysics, University of California, San Diego, La Jolla, CA 92093, USA}

\author{Carlo~Baccigalupi~\orcidlink{0000-0002-8211-1630}}
\affiliation{The International School for Advanced Studies (SISSA), via Bonomea 265, I-34136 Trieste, Italy}
\affiliation{The National Institute for Nuclear Physics (INFN), via Valerio 2, I-34127, Trieste, Italy}
\affiliation{The Institute for Fundamental Physics of the Universe (IFPU), Via Beirut 2, I-34151, Trieste, Italy}

\author{Darcy~R.~Barron~\orcidlink{0000-0002-1623-5651}}
\affiliation{Department of Physics and Astronomy, University of New Mexico, Albuquerque, NM 87131 USA}

\author{Bryce~Bixler}
\affiliation{Department of Physics, University of California, San Diego, La Jolla CA 92093, USA}

\author{Yuji~Chinone~\orcidlink{0000-0002-3266-857X}}
\affiliation{International Center for Quantum-field Measurement Systems for Studies of the Universe and Particles (QUP), High Energy Accelerator Research Organization (KEK), Tsukuba, Ibaraki 305-0801, Japan}
\affiliation{Kavli IPMU (WPI), UTIAS, The University of Tokyo, Kashiwa, Chiba 277-8583, Japan}

\author{Matthew~R.~Chu~\orcidlink{0000-0001-5884-911X}}
\affiliation{Department of Physics, University of California, San Diego, La Jolla CA 92093, USA}

\author{Kevin~T.~Crowley~\orcidlink{0000-0001-5068-1295}}
\affiliation{Department of Astronomy \& Astrophysics, University of California, San Diego, La Jolla, CA 92093, USA}

\author{Nicole~Farias~\orcidlink{0000-0002-4101-2513}}
\affiliation{Department of Physics, University of California, Berkeley, CA 94720, USA}

\author{Takuro~Fujino~\orcidlink{0000-0002-1211-7850}}
\affiliation{International Center for Quantum-field Measurement Systems for Studies of the Universe and Particles (QUP), High Energy Accelerator Research Organization (KEK), Tsukuba, Ibaraki 305-0801, Japan}

\author{Masaya~Hasegawa~\orcidlink{0000-0003-1443-1082}}
\affiliation{International Center for Quantum-field Measurement Systems for Studies of the Universe and Particles (QUP), High Energy Accelerator Research Organization (KEK), Tsukuba, Ibaraki 305-0801, Japan}
\affiliation{Institute of Particle and Nuclear Studies (IPNS), High Energy Accelerator Research Organization (KEK), Tsukuba, Ibaraki 305-0801, Japan}

\author{Masashi~Hazumi~\orcidlink{0000-0001-6830-8309}}
\affiliation{Institute of Particle and Nuclear Studies (IPNS), High Energy Accelerator Research Organization (KEK), Tsukuba, Ibaraki 305-0801, Japan}
\affiliation{Department of Physics and Center for High Energy and High Field Physics (CHiP), National Central University, Taoyuan City, Taiwan}

\author{Haruaki~Hirose~\orcidlink{0000-0003-3336-134X}}
\affiliation{Institute of Particle and Nuclear Studies (IPNS), High Energy Accelerator Research Organization (KEK), Tsukuba, Ibaraki 305-0801, Japan}
\affiliation{Department of Physics, Graduate School of Engineering Science, Yokohama National University, Yokohama, Kanagawa 240-8501, Japan}

\author{Jennifer~Ito}
\affiliation{Westmont College, Santa Barbara, CA 93108, USA}

\author{Oliver~Jeong~\orcidlink{0000-0001-5893-7697}}
\email[Corresponding author: \href{mailto: objeong@lbl.gov}{objeong@lbl.gov}]{}
\affiliation{CNRS-UCB International Research Laboratory, Centre Pierre Binétruy, IRL 2007, CPB-IN2P3, Berkeley, CA 94720, USA}
\affiliation{Physics Division, Lawrence Berkeley National Laboratory, Berkeley, CA 94720, USA}

\author{Daisuke~Kaneko~\orcidlink{0000-0003-3917-086X}}
\affiliation{International Center for Quantum-field Measurement Systems for Studies of the Universe and Particles (QUP), High Energy Accelerator Research Organization (KEK), Tsukuba, Ibaraki 305-0801, Japan}

\author{Brian~Keating}
\affiliation{Department of Physics, University of California, San Diego, La Jolla CA 92093, USA}

\author{Akito~Kusaka~\orcidlink{0009-0004-9631-2451}}
\affiliation{Physics Division, Lawrence Berkeley National Laboratory, Berkeley, CA 94720, USA}
\affiliation{Department of Physics, The University of Tokyo, Tokyo 113-0033, Japan}
\affiliation{Kavli Institute for the Physics and Mathematics of the Universe (WPI), UTIAS, The University of Tokyo, Kashiwa, Chiba 277-8583, Japan}
\affiliation{Research Center for the Early Universe, School of Science, The University of Tokyo, Tokyo, 113-0033, Japan}

\author{Adrian~T.~Lee~\orcidlink{0000-0003-3106-3218}}
\affiliation{Department of Physics, University of California, Berkeley, CA 94720, USA}
\affiliation{Physics Division, Lawrence Berkeley National Laboratory, Berkeley, CA 94720, USA}

\author{Masaaki~Murata~\orcidlink{0000-0003-4394-4645}}
\email[Corresponding author: \href{mailto: mmurata@cmb.phys.s.u-tokyo.ac.jp}{mmurata@cmb.phys.s.u-tokyo.ac.jp}]{}
\affiliation{Department of Physics, The University of Tokyo, Tokyo 113-0033, Japan}

\author{Lucio~Piccirillo~\orcidlink{0000-0001-7868-0841}}
\affiliation{Jodrell Bank Centre for Astrophysics, University of Manchester, Manchester, UK}

\author{Christian~L.~Reichardt~\orcidlink{0000-0003-2226-9169}}
\affiliation{School of Physics, The University of Melbourne, Parkville, VIC 3010, Australia}

\author{Kana~Sakaguri~\orcidlink{0000-0001-5667-8118}}
\affiliation{Department of Physics, The University of Tokyo, Tokyo 113-0033, Japan}

\author{Praween~Siritanasak~\orcidlink{0000-0001-6830-1537}}
\affiliation{National Astronomical Research Institute of Thailand, Chiangmai 50180, Thailand}

\author{Satoru~Takakura~\orcidlink{0000-0001-9461-7519}}
\affiliation{Department of Physics, The University of Tokyo, Tokyo 113-0033, Japan}

\author{Sayuri~Takatori~\orcidlink{0000-0002-8705-9624}}
\affiliation{Research Institute for Interdisciplinary Science (RIIS), Okayama University, Okayama 700-8530, Japan}

\author{Osamu~Tajima~\orcidlink{0000-0003-2439-2611}}
\affiliation{International Center for Quantum-field Measurement Systems for Studies of the Universe and Particles (QUP), High Energy Accelerator Research Organization (KEK), Tsukuba, Ibaraki 305-0801, Japan}
\affiliation{Department of Physics, Kyoto University, Kyoto 606-8502, Japan}

\author{Kyohei~Yamada~\orcidlink{0000-0003-0221-2130}}
\affiliation{Joseph Henry Laboratories of Physics, Jadwin Hall, Princeton University, Princeton, New Jersey 08544, USA}

\author{Yuyang~Zhou~\orcidlink{0000-0002-5878-4237}}
\affiliation{Department of Physics, University of California, Berkeley, CA 94720, USA}

\collaboration{The \textsc{Polarbear} collaboration}

\date{\today}

\begin{abstract}
We present a search for polarization oscillation of the Crab Nebula, also known as Tau A, at millimeter wavelengths using observations with the Simons Array, the successor experiment to \textsc{Polarbear}. We follow up on previous work by \textsc{Polarbear} using 90 GHz band data of the 2023 observing season of the Simons Array to evaluate the variability of Tau A's polarization angle. Tau A is widely used as a polarization angle calibration source in millimeter-wave astronomy, and thus it is necessary to validate the stability. Additionally, an interesting application of the time-resolved polarimetry of Tau A is to search for axion-like particles (ALPs). We do not detect a global signal across the frequencies considered in this analysis and place a median 95\% upper bound of polarization oscillation amplitude $A<0.12^{\circ}$ over oscillation frequencies from 3.39 year$^{-1}$ to 1.50 day$^{-1}$. This constrains the ALP-photon coupling at a median 95\% upper bound of $g_{a\gamma\gamma}< 3.84\times 10^{-12}\times\left(m_a/10^{-21}\,\mathrm{eV}\right)$ in the mass range from $4.4\times10^{-22}$ to $7.2\times10^{-20}$ eV, assuming the ALP constitutes all of dark matter, its field is a stochastic Gaussian field, and it is the sole source of Tau A's polarization angle oscillation. Additionally, we do not detect signal at the frequencies where 2.5$\sigma$ hints were previously reported by \textsc{Polarbear}, but we do not exclude these signals at the 95\% confidence level.
\end{abstract}

\maketitle

\section{\label{sec:intro} Introduction}
Tau A, also known as the Crab Nebula, is a supernova remnant whose polarization emission is widely used as a polarization angle calibration source in millimeter-wave astronomy~\cite{aumont1,aumont2}. The \textsc{Polarbear} collaboration recently demonstrated a method of investigating the time-varying polarization angle of Tau A using its 150 GHz band observations and reported a 2.5$\sigma$ hint of polarization oscillation at 1/61 day$^{-1}$ and 1/52 day$^{-1}$~[\cite{PB24}, hereafter \citetalias{PB24}]. This potential variation is crucial for understanding the internal dynamics of Tau A and may be treated as a systematic error when using Tau A as a calibration source. Here we extend the results by applying the method to data from independent observations at 90 GHz with \textsc{Polarbear}-2a (PB-2a), the first receiver of the successor experiment of \textsc{Polarbear}, the Simons Array. 
\par
In addition to probing the stability of Tau A as a millimeter-wave calibration source,~\citetalias{PB24} explored its applications to the search for axion-like particles (ALPs), assuming the observed time-varying polarization angle is induced by electromagnetic coupling to ALPs. The reported hint of signal is inconsistent with other studies searching for ALP-induced oscillations in the same mass range~\cite{spt3g_axion, fedderke, PPTA, quasarh1821}, highlighting the need for this paper to apply \citetalias{PB24}'s method as a consistency check. Originally proposed to solve the strong CP problem \cite{weinberg_phs_rev_lett, peccei_quinn_phys_rev_d, peccei_quinn_phys_rev_lett, wilczek_phys_rev_let}, the QCD axion and its generic pseudoscalar form, the ALP, are prime particle candidates of dark matter and are therefore of broad astrophysical interest. The ALP does not solve the strong CP problem and is motivated by string theories~\cite{weinberg_phs_rev_lett, arvanitaki, cicoli}. Without a fixed relationship between its mass and the Standard Model, the ALP mass is theoretically unconstrained over a broad range \cite{ferreira, hui}. For masses around 10$^{-22}$ eV, ALP dark matter resolves existing small-scale tensions in the standard cold dark matter model and fits the observed phenomenology of dark matter \cite{hu_phys_rev_lett, hui_phys_rev_d}.
\par
In this paper, we exploit the ALP's coupling to electromagnetism through the Chern-Simons term, introducing a birefringence of the ALP field \cite{carrol_field_roman, harari} and rotating the polarization angle of the photon propagating through the ALP field as
\begin{equation}
\psi=\frac{g_{a\gamma\gamma}}{2}(\phi_{\mathrm{obs}}-\phi_{\mathrm{emit}})
\label{eq:polangle}
\end{equation}
where $g_{a\gamma\gamma}$ is the ALP-photon coupling, $\phi_{\mathrm{obs}}$ is the ALP field where the photon is observed, and $\phi_{\mathrm{emit}}$ is the ALP field where the photon was emitted. We consider the case in which the ALP behaves as wave-like dark matter, for which the time variation of its field is described by a single mode sinusoid with period $h/(m_ac^2)$ where $m_a$ is the ALP mass. For the mass range investigated in this study, $4.4\times10^{-22}$ to $7.2\times10^{-20}$ eV, the ALP's coherence length is much shorter than the distance to Tau A \cite{kaplan} and the ALP Compton wavelength is much smaller than the size of Tau A \cite{hester}. Thus, as in \citetalias{PB24}, the ALP-induced polarization oscillation of Tau A can be modeled as

\begin{equation}
\psi (t)=\psi_0 + \frac{g_{a\gamma\gamma}\phi_0}{2}\sin\left(\frac{m_ac^2}{\hbar} t+\theta\right),
\label{eq:axioninduced_signal}
\end{equation}
where $\psi_0$ is the polarization angle of Tau A.
\par
In addition to studying the polarization angle variability of Tau A as a polarization angle calibrator at 90 GHz to extend \textsc{Polarbear}'s study at 150 GHz, using the 90 GHz band data of PB-2a to study axion-like polarization angle oscillations enables a check against frequency-dependent systematics such as Faraday rotations to \textsc{Polarbear}'s 150 GHz ALP search. This investigation into the PB24 ALP search is feasible because the ALP signal coherence time is greater than 100 years for the mass range of interest of this paper and thus remains coherent across \textsc{Polarbear}'s observations from 2016–2017 and PB-2a from 2023. To follow up on the results of the previous publication, the analysis method is purposely kept as identical as possible.
\par
This paper is organized as follows: In Sec.~\ref{sec:instrument}, we provide an overview of the instrument and its polarization modulator. In Sec~\ref{sec:dataset}, we describe the Tau A observations used in this analysis. In Sec.~\ref{sec:analysis}, we describe the analysis method used in this paper, closely adapted from \citetalias{PB24}, and in Sec.~\ref{sec:data_validation}, the framework used to check for internal consistency against spurious results. In Sec.~\ref{sec:results}, we present the results of the analysis in Tau A polarization angle oscillation and limits on the ALP-photon coupling constant. We close in Sec.~\ref{sec:conclusion} with concluding remarks and expectations of a follow-up study by the Simons Array.

\section{\label{sec:instrument}Instrument Overview}
\subsection{Site, telescope and receiver}
\begin{figure}
\centering
\includegraphics[width=\columnwidth]{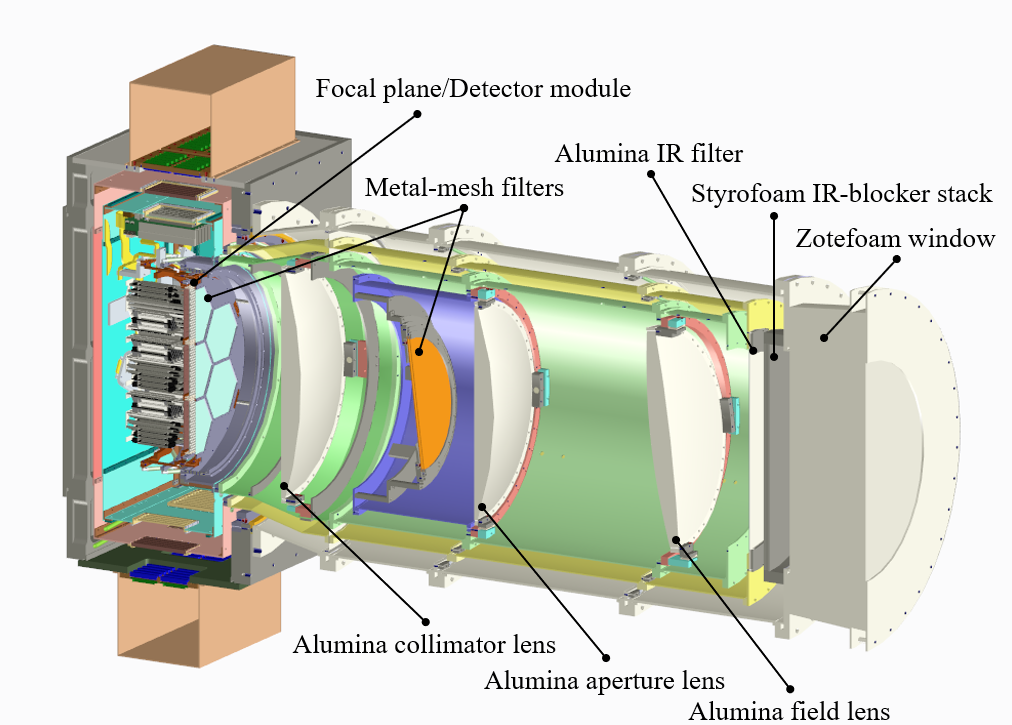}
\caption{\label{fig:receiver}Cutaway view of the PB-2a receiver cryostat, with key elements labeled.}
\end{figure}

The Simons Array is a millimeter-wave observatory on Cerro Toco in the Atacama Desert of Chile at an elevation of 5,200 m \cite{Suzuki2016-yw}. It upgrades and expands the \textsc{Polarbear} instrument into an array consisting of two dichroic polarimeters: \textsc{Polarbear}-2a (PB-2a) and PB-2b. Each telescope consists of a 2.5 m primary reflector arranged in an off-axis Gregorian-Dragone configuration that satisfies the Mizuguchi-Dragone condition, minimizing cross-polarization systematics and aberrations \cite{Mizugutch2005-gw,Dragone1978-jo,Tran2008-ro,Takakura2018-dt_MD_breaking}. The observation bands of PB-2a are centered at 90 and 150 GHz, with beams of 5.2 and 3.5 arcmin FWHM, respectively \cite{Suzuki2014-SA_dets, Westbrook2016-SA_dets}. The telescope houses a receiver cryostat (see Fig.~\ref{fig:receiver}) containing a 500 mm Zotefoam vacuum window, three alumina re-imaging lenses at 4 K, a Lyot stop at 4 K, and a 365 mm focal plane at 0.3 K with more than 7,600 transition-edge sensor detectors. A hemispherical lenslet focuses the output of the re-imaging lenses onto a detector pixel on the focal plane \cite{Siritanasak2016-bw}. Each pixel utilizes a broadband sinuous antenna to couple optical signals of two orthogonal polarization states into a microstrip circuit. The circuit separates the signals into the two PB-2a frequency bands using band-pass filters, which then route the signals to the detectors, yielding four detectors per pixel.

\subsection{Half-wave plate}
PB-2a utilizes a continuously rotating half-wave plate (HWP) as a polarization modulator, rotating at 2 Hz and placed immediately in front of the receiver window at ambient temperature \cite{Hill2016}. The HWP up-converts the incident polarization signal above $1/f$ unpolarized fluctuations, thereby suppressing low-frequency noise. Given Stokes parameters \textit{I}, \textit{Q}, and \textit{U}, the raw time-ordered data (TOD) of a detector is modeled as

\begin{equation}
\begin{aligned}
d_{m}(t) = I(t) + \varepsilon \mathrm{Re}\left\{ \left[ Q(t) + i U(t) \right]e^{-i\left(4\chi\left(t \right) + 2\theta_{\mathrm{det}} \right)} \right\}\\
+ A\left(\chi,t \right) + \EuScript{N}_{m},
\label{eq:modulated_tod}
\end{aligned}
\end{equation}
where $\varepsilon$ is the polarization efficiency, $\chi\left(t \right)$ is the HWP rotation angle, $\theta_{\mathrm{det}}$ is the angle of the detector in telescope coordinates, $A\left(\chi,t \right)$ is the HWP synchronous signal, and $\EuScript{N}_{m}$ is the white noise of the detector. Note that polarization is modulated at four times the HWP frequency as captured in the modulation function, $e^{-i4\chi\left(t \right)}$.
The HWP synchronous signal is generated by many mechanisms related to instrumental polarization and non-idealities of the HWP. In a simplified model, the HWP synchronous signal is described by
\begin{equation}
\begin{aligned}
A\left(\chi,t \right) = \sum_{n=1}\mathrm{Re}\left\{A_n(t)e^{-in\chi(t)}\right\},
\label{eq:hwpss}
\end{aligned}
\end{equation}
where $A_n$ is the slowly-varying amplitude of the $n$th harmonic. Throughout this paper, we refer to the 2nd and 4th harmonics of the HWP synchronous signal as the 2f and 4f signals, respectively. While an estimate of the HWP synchronous signal is subtracted from the TODs in the filtering process, it is also a crucial tool in the calibration process as discussed in Section~\ref{sec:calibration}.
\section{\label{sec:dataset}Dataset}
\subsection{Observations\label{Observations}}
In the observing season from September to December 2023, PB-2a observed Tau A on a near-daily schedule, with 4 separate observations each day. While both 90 and 150 GHz data are available from these observations, the focus of this analysis is placed on the 90 GHz dataset, and we leave the larger dataset to be analyzed for a future publication. Given that the PB-2a 90 GHz beam with $5.2'$ FWHM is larger than the Tau A angular size of $7'\times5'$~\cite{Green}, the internal structure of Tau A is not resolved, but rather averaged during the calculation of the Tau A polarization angle (Sec~\ref{sec:statunc}).
\par
Tau A is observed using a raster scan pattern, with the telescope continuously tracking Tau A in elevation while scanning back-and-forth in azimuth within 8$^{\circ}$ at 0.4$^{\circ}$/s. Between scans in azimuth, the elevation is stepped to ensure the entire focal plane observes Tau A. In contrast with \citetalias{PB24} which only observed Tau A as it set from 39$^{\circ}$ to 30$^{\circ}$ in elevation, PB-2a observed Tau A for the entirety of its visibility as it rose and set between 30$^{\circ}$ and 45$^{\circ}$. Throughout a day of observations, it observed Tau A in 4 different scan-types of different elevation: rising (30$^{\circ}$ -- 37$^{\circ}$), middle 1 (37$^{\circ}$ -- 45$^{\circ}$), middle 2 (45$^{\circ}$ -- 37$^{\circ}$), and setting (37$^{\circ}$ -- 30$^{\circ}$). The typical duration of each observation was 82 minutes. The observation efficiency is summarized in Table~\ref{tab:obs_eff}.

\subsection{Data Selection}
We apply similar data selection criteria as in \citetalias{PB24}. The data selection efficiency is summarized in Table~\ref{tab:data_selection}. Note that the detector selection efficiency is normalized to 1088 detectors, which is the number of 90 GHz detectors with valid calibration.

There are a few data selection criteria that differ from the \citetalias{PB24} analysis. In terms of the observation data selection, we cut observations that do not follow the pre-defined scan-types listed in Sec.~\ref{Observations}. We are left with 76 observations of Tau A that pass our data selection. For these observations, we make unpolarized intensity maps of Tau A and calculate the signal-to-noise ratio (SNR) of these intensity maps for each detector with valid gain calibrations. We calculate the signal and noise in map-space using the same integration regions we use for polarization angle estimation of the coadded maps as described in Sec.~\ref{sec:statunc}. We cut detectors that observe Tau A intensity with SNR $<$ 10. Additionally, these intensity maps are used for observation-by-observation pointing calibration as is described in Sec.~\ref{sec:pointing_cal}.

\begin{table}[b]
\caption{\label{tab:obs_eff}
Observation efficiency.
}
\begin{ruledtabular}
\begin{tabular}{lc}
\textrm{Observation breakdown}&
\textrm{}\\
\colrule
from & September 1, 2023\\
until & December 27, 2023\\
Total calendar time & 2,832 hr\\
Total Tau A visibility time & 640 hr\\
\\
\textit{Observation yield} & \\
Telescope issues & 77.8\%\\
Moon avoidance & 90.4\%\\
Fridge \& detector tuning issues & 93.2\%\\
HWP issues & 98.1\%\\
Operator error & 98.5\%\\
\\
Total time observing Tau A & 371 hr\\
\\
Cumulative observation efficiency & 57.9\%\\
\end{tabular}
\end{ruledtabular}
\end{table}

\bgroup
\def\arraystretch{1.2}
\begin{table}[b]
\caption{\label{tab:data_selection}
Data selection efficiency.
}
\begin{ruledtabular}
\begin{tabular}{lr}
\textrm{Type of selection}&
\textrm{Selection efficiency}\\
\colrule
\textit{Observation selection} & \\
Instrumental issues & 56.7\%\\
Within specified scan regions & 85.5\% \\
PWV conditions & 75.0\% \\
Angular distance to sun/moon & 92.9\% \\
Focal plane temperature & 97.4\% \\
\\
\textit{Detector selection} &  \\
Detector power spectral density & 69.8\% \\
Tau A intensity SNR & 89.5\% \\
HWP synchronous signal phases & 99.3\%\\
\\
\textit{Subscan selection} & \\
Common mode glitches & 99.9\% \\
Common mode power spectral density & 99.5\% \\
Individual detector glitch & 99.9\% \\
\\
Cumulative data selection & 20.3\%

\end{tabular}
\end{ruledtabular}
\end{table}
\egroup{}

\section{\label{sec:analysis}Analysis Method}
\subsection{TOD processing and demodulation}\label{tod_processing}

The first step in extracting the intensity and polarization signals from the raw TODs is to subtract the HWP synchronous signal. We reconstruct the angle of the HWP using data from an optical encoder, which we then downsample to the timestamps of our detector TODs. For the first eight harmonics of the HWP rotation frequency, we bandpass filter the TOD at the harmonic with a bandwidth of 0.2 Hz and demodulate the TOD with respect to that harmonic. The result is a template for the time-varying amplitude of the harmonic, which we subtract from the TOD. After subtracting the HWP synchronous signal, the polarization signal can be demodulated and extracted by multiplying $d_{m}\left(t\right)$ by twice the conjugate of the modulation function, $2e^{i4\chi\left(t\right)}$, and low-pass filtering. To extract the intensity signal, we low-pass filter $d_{m}\left(t\right)$ after subtracting the HWP synchronous signal. In both cases, the low-pass filter has a cutoff frequency of 3.8 Hz.

There is leakage of the intensity signal into the polarization signal (I2P) due to detector non-linearity and instrumental polarization \cite{satoru_pb_whwp}. We use the extracted signals to estimate and remove this leakage using a principal component analysis (PCA), following the approach utilized in \cite{PB_large_patch}. We compute the covariance matrices between the $I$ and $Q(U)$ TODs during the entire observation while masking out the regions where the telescope is not scanning with constant velocity and where the detector pointing is within a $12'$ radius centered on Tau A. Assuming the intensity signal is dominated by unpolarized low-frequency atmospheric fluctuations, we perform PCA on these covariance matrices in order to estimate the leakage coefficients following \cite{satoru_pb_whwp}. After subtracting the estimated leakage, we fit the timestream of each constant-velocity subscan with a first-order polynomial and subtract it. When fitting the polynomial, we mask out sections of the TOD when the detector pointing is within a $12'$ radius centered around Tau A.

We note that this method of leakage estimation differs from that used in \citetalias{PB24} as we estimate leakage coefficients for each detector on an observation-by-observation basis rather than using fixed leakage coefficients for each detector. We will refer to the observation-by-observation leakage estimation method as the variable leakage model, and the method used in \citetalias{PB24} as the fixed leakage model. We adopted the variable leakage model through iterative data validation tests. Both the fixed and variable leakage models passed the data validation tests, but the fixed leakage model required excluding five additional observations. These five observations have statistically significant variation exclusively in the I2U leakage. This variation does not correlate with any studied observing or instrumental conditions, such as the precipitable water vapor (PWV) or the bias conditions of our detectors. We concluded that the variable leakage model does not lead to a systematic misestimation of leakage and captures this variation well. Further discussion of possible systematic uncertainties due to this leakage estimation model is in Sec.~\ref{I2P_systematic}.

\subsection{\label{sec:calibration}Calibration}

\subsubsection{\label{sec:reldetgain}Relative Detector Gains}
We use multiple observations of Jupiter to calibrate a chopped thermal source located behind the secondary mirror \cite{Kaneko2024_stim}. This thermal source is then used to calibrate the amplitude of the 2f component of the HWP synchronous signal in units of $\mathrm{K}_{\mathrm{CMB}}$. We use the 2f signal because it is the brightest HWP synchronous signal harmonic. We expect that the 2f signal exists due to differential transmission and absorption between the crystal axes in the HWP \cite{kusaka_RSI}, and so the amplitude of the 2f signal varies linearly with the brightness of the atmosphere. We fit a linear model between the 2f amplitude in $\mathrm{K}_{\mathrm{CMB}}$ and our proxy for atmospheric brightness, $\mathrm{PWV}/\sin\left(\mathrm{elevation}\right)$, for each detector. We use this model to estimate the expected 2f brightness for a given observation, which is then used for gain calibration. We note here that because this analysis is focused on estimating the polarization angle of Tau A, we do not apply an overall gain calibration to our instrument coadded data that accounts for effects such as imperfect polarization efficiency in the PB-2a instrument.

\subsubsection{\label{sec:time_const}Time Constants}
The time constants of our detectors cause an apparent phase lag between the reconstructed angle of the HWP and the detector TOD, appearing as an error in the polarization angle, which must be calibrated. Assuming that any measured variation in the phase of the 2f signal is due to the effect of the time constants and a single pole transfer function for the time constants, the phase rotation at 2f is not identical to the phase rotation at 4f, where the Tau A polarization signal lives. For each detector, we measure:

\begin{equation}
\delta \psi^{2f(4f)}_{t,j} = \mathrm{average}_{j}\left( \psi^{2f(4f)}_{t,j}\right) - \psi_{t,j}^{2f(4f)}
\label{eq:delta_phase_2f4f}
\end{equation}
where $\psi^{2f(4f)}$ is the phase of the 2f(4f) signal, $t$ is the index of the observation, $j$ is the index of the detector, and $\mathrm{average}_j(...)$ is an average over all observations. We fit a linear relationship between $\delta \psi_{t,j}^{2f}$ and $\delta \psi_{t,j}^{4f}$ as $\delta \psi^{4f}_{t,j} = C^{\tau}_{j}\delta \psi^{2f}_{t,j}$ and use this model to rotate the polarization basis at 4f using the measured variation in 2f phase for a given observation and detector:

\begin{equation}
Q_{t,j}' + iU_{t,j}' = \left(Q_{t,j} + i U_{t,j}\right)\exp\left(-iC_{j}^{\tau}\delta\psi_{t,j}^{2f}\right).
\label{eq:phase_cor_2f}
\end{equation}
Here, $Q_{t,j}$ and $U_{t,j}$ are the TODs for the Stokes parameters after the demodulation procedure described in Sec.~\ref{tod_processing}. We note that a unique $C^{\tau}_{j}$ is fit for each individual detector in our analysis, as we expect this factor to vary based on the mean time constant of a given detector.

\subsubsection{\label{sec:pointing_cal}Pointing}
We calibrate the pointing for each detector for each observation using the Tau A intensity signal. For each detector in a given observation, we fit the measured intensity signal to an elliptical Gaussian model in equatorial coordinates. The offsets from the centers of the Gaussian fits to the expected Tau A position are then applied to the detector pointing. We discuss possible errors associated with our pointing reconstruction in Sec. \ref{pointing_systematic}.

\subsubsection{\label{sec:pol_angle_cal}Detector polarization angles}
We use the full season of Tau A observations to calibrate the polarization angles of our detectors. Nominally, this means constructing season co-added maps of Tau A for each detector, $j$, and rotating $Q_{t,j} + iU_{t,j}$ for each observation, $t$, by the offset between the measured Tau A polarization angle in the co-added maps for each detector and the Tau A polarization angle as measured by IRAM \cite{aumont1}.

When applying the polarization angle calibration for the entire season of data, a systematic correlation between the azimuth and elevation of an observation and the extracted Tau A polarization angle was discovered. In order to prevent any systematic contamination, we instead calibrate the polarization angles independently for the four distinct observation types described in Sec. \ref{Observations}. Therefore, instead of constructing co-added maps over the course of the entire observing season, we construct co-added maps for each observation type and apply the polarization angle calibrations independently. This leads to a loss of sensitivity for some modes, and we simulate this effect in our transfer function, which we show in Fig. \ref{fig:transferfunction}.

\subsubsection{Relative polarization angles between observations}
In order to avoid systematic misestimation of the measured polarization angle between observations, we make use of the HWP synchronous signal. We assume that the 4f signal is dominated by optical systematics \cite{satoru_pb_whwp} and that the phase of this signal remains constant in instrumental coordinates between observations. We first attempt to correct for the phase lag due to the detector time constants as in Sec. \ref{sec:time_const}. The 4f phase variation for an individual detector for a given observation after this correction is

\begin{align}
\tilde{\delta\psi}^{4f}_{t,j} &= \delta \psi_{t,j}^{4f} - C^{\tau}_{j}\delta\psi_{t,j}^{2f}.
\end{align}
We expect that any systematic rotations common to measured polarization angles in the focal plane, such as a misestimation of the HWP angle or a timing offset, will shift the distribution of $\tilde{\delta{\psi}}_{t,j}^{4f}$ away from $0$. We estimate this systematic rotation as the median of the 4f variation over all detectors

\begin{align}
\alpha_{t} &\equiv \mathrm{median}_{j}\left(\tilde{\delta\psi}^{4f}_{t,j} \right),
\label{eq:cor_4f_phase}
\end{align}
and correct for this offset by rotating the polarization basis of our timestreams as

\begin{equation}
Q''_{t,j} + iU''_{t,j} = \left(Q'_{t,j} + i U'_{t,j}\right)\exp\left(-i\alpha_{t}\right).
\label{eq:med_4f_phase_cor}
\end{equation}

\subsection{\label{sec:statunc}Angle measurement and statistical uncertainty}

\begin{figure}
\includegraphics[width=\columnwidth]{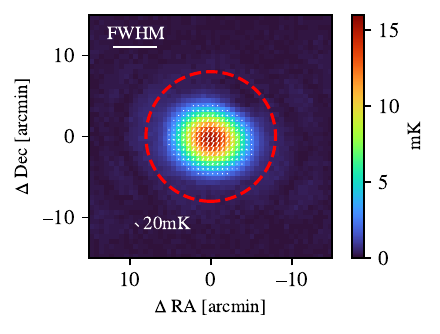}
\caption{\label{season_coadd}Season co-add polarization intensity map of Tau A. The white lines represent the orientation of the polarization angle in each pixel. The red dashed circle covers the integration area used to estimate the polarization angle for individual observations and to evaluate systematic errors. The FWHM of the PB-2a beam is shown for comparison.}
\end{figure}

Our mapmaking pipeline projects the filtered and calibrated TODs onto flat sky maps with pixel widths of $2.4'$ using a cylindrical equal-area projection. We estimate the polarization angle of Tau A in map-space following the method outlined in \citetalias{PB24}

\begin{equation}
    d_{t} = \frac{1}{2}\textrm{tan}^{-1}\left( \frac{\sum_{p}^{\diameter<16'} U_{p}^{\textrm{Tau\ A}}}{\sum_{p}^{\diameter<16'} Q_{p}^{\textrm{Tau\ A}}}\right)
\end{equation}
where $d_t$ is the measured polarization angle for observation $t$, $Q_p$ and $U_p$ are the $p$-th $Q$ and $U$ map pixel in the observation co-added map. We show the co-added map of all observations in this analysis in Fig.~\ref{season_coadd} and highlight the integration area used to estimate the polarization angle. We note that we have increased the diameter of the integration area to $16'$ to account for the increased beam size in the PB-2a experiment ($5.2'$) compared to the \textsc{Polarbear} experiment ($3.6'$). We also follow the bootstrap method outlined in \citetalias{PB24} to generate white noise realizations and use these to estimate the statistical uncertainty on our polarization angle estimate, $\sigma_{t}$. Our median statistical uncertainty in the polarization angle of Tau A is $0.25^\circ$ per observation.

\subsection{Polarization oscillation spectrum estimation}
For consistency, and to compare our results more directly with the \citetalias{PB24} results, we choose to report on the same frequencies as the finer spaced frequencies in their analysis. The total length of the observing season considered in this analysis is shorter, so we cut out the lowest frequency modes to which we are insensitive. This leaves us with the following frequencies:
\begin{equation}
    f_{\mathrm{bins}} = \left\{ \frac{45}{4860\Delta T}, \frac{46}{4860\Delta T},...,\frac{2420}{4860\Delta T}\right\}
\end{equation}
where $\Delta T$ corresponds to one sidereal day ($\Delta T \approx 1 - 1/366.24\ $ days). Our near-daily observing schedule means that power in frequencies below $1/2\Delta T$ is degenerate with frequencies between $1/2\Delta T$ and $3/2\Delta T$ due to aliasing. Through simulations, we confirmed that a signal with frequency between $1/2\Delta T$ and $3/2\Delta T$ will be measured at approximately 90\% amplitude at the aliased frequency below $1/2\Delta T$. Therefore, we extend the scientific results to 
\begin{equation}
\label{eq:frequencies}
    f_\mathrm{bins,result} = \left\{ \frac{N}{2\Delta T}+f_{\mathrm{bins}}\bigg| N=0,1,2\right\}.
\end{equation}
We confirm that the data is internally consistent at the extended frequencies by scaling the systematic estimates and null statistics described in Sec. \ref{sec:data_validation} by this aliasing factor and find that we still pass our data validation criteria.

\begin{figure}
\includegraphics[width=\columnwidth]{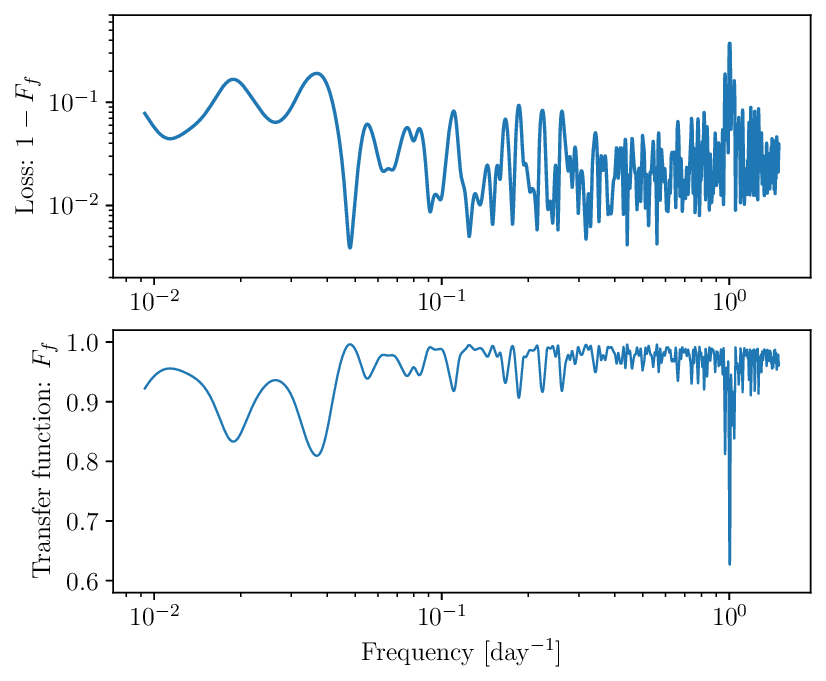}
\caption{\label{fig:transferfunction}Transfer function $F_f$ (bottom) and loss $1-F_f$ (top) for the frequency bins, described by Eq.~\ref{eq:frequencies}. The transfer function is averaged over all the phases of oscillations.}
\end{figure}

We follow the exact procedure outlined in \citetalias{PB24} to estimate the polarization oscillation amplitude at each of the frequencies. To summarize briefly, we perform least-squares spectral analysis (LSSA) on our timestream of Tau A polarization angles and their associated uncertainties. Similarly to \citetalias{PB24}, our estimate of the signal amplitude is biased, and we quantify this bias by estimating our transfer function, $F_{f}$, in a manner identical to the procedure outlined in \citetalias{PB24}. The method for polarization angle calibration differs between the two analyses, which we take into account when calculating the transfer function for this analysis. The full transfer function can be seen in Fig.~\ref{fig:transferfunction}.

We use an identical test statistic as \citetalias{PB24} to quantify the global significance for the preference of a signal:
\begin{align}
    \Delta \chi^{2}_{f} &= \sum_{m,n\in\{\mathrm{real},\mathrm{imag}\}}\tilde{d}_{f}^{m}\tilde{N}_{mn}^{-1}\tilde{d}_{f}^{n}\\
    \Delta \chi^{2}&=\textrm{max}_f\left( \Delta \chi^{2}_{f}\right),
\end{align}
where $d_{f}$ is the complex LSSA amplitude of the signal at frequency $f$ and $\tilde{N}_{mn}$ is our noise covariance matrix estimated using the same noise realizations we use to determine our statistical uncertainty. To follow up on the hint of signal seen in \citetalias{PB24}, we additionally use $\Delta \chi^{2}_{f}$ to test for the local significance of a signal at the two frequencies reported on in \citetalias{PB24} (hereafter, the two PB24 frequencies): $1/61.4\ \textrm{day}^{-1},\ 1/52.1\ \textrm{day}^{-1}$.

\section{\label{sec:data_validation}Data Validation}
We utilize a blind analysis and only unblind the estimated polarization angles of Tau A and the associated oscillation spectrum after checking for internal consistency in our data using null tests. In this section, we will summarize the null tests and their results. We also report on the impact of potential sources of systematic error and utilize the weighted-correlation method to search for additional systematic contamination.

\subsection{Null test}

In constructing the null tests for probing systematic errors across the entire frequency spectrum, we follow the analysis presented in \citetalias{PB24}. We split the data either based on a characteristic of an individual observation or on individual detectors and check for consistency between the two halves. Appendix~\ref{apx:null_test_splits} provides a brief description for each observation and detector split. Following \citetalias{PB24}, we calculate $\chi_{\mathrm{null}}^{\mathrm{DC}}, \chi^{\mathrm{AC}}_{\mathrm{null}}$ and $\chi_{\mathrm{null}}^{2}$. $\chi_{\mathrm{null}}^{\mathrm{DC}}$ represents the difference in the noise-weighted mean polarization angle between each half of the split; $\chi^{\mathrm{AC}}_{\mathrm{null}}$ and $\chi_{\mathrm{null}}^{2}$ capture the noise-weighted difference in the real/imaginary LSSA amplitudes in each half of the split. We use these statistics to compute six representative null statistics for both the observation-based and detector-based splits. We calculate the probability-to-exceed (PTE) for these representative null statistics by comparing to 1,440 bootstrap noise simulations. We then use these PTE to check whether our pass criteria are met:

\begin{enumerate}
    \item The PTE of the lowest PTE value\footnote{As in \citetalias{PB24}, the PTE of the lowest (highest) PTE value is obtained by counting the number of simulations whose lowest (highest) PTE value is smaller (larger) than the observed lowest (highest) PTE value.} is greater than 0.05.
    \item The PTE of the highest PTE value is greater than 0.05.
\end{enumerate}
Note that we require the pass criteria to be met for both the observation-based and detector-based null tests separately.

We also include null tests intended to probe for systematic errors at the two PB24 frequencies: $1/61.4$ day$^{-1}$ and $1/51.2$ day$^{-1}$. We choose three representative null statistics to probe these frequencies: (1) the average $\chi_{\mathrm{null}}^{\mathrm{AC}}$ over both frequencies and all data splits; (2) the total $\chi^{2}_{\mathrm{null}}$ over both frequencies and all data splits; (3) the most extreme $\chi_{\mathrm{null}}^{2}$ over both frequencies and all data splits. The pass criteria are identical to those used for the full spectrum null tests.

The PTEs of the representative null statistics for both the full spectrum null tests and the two PB24 frequency null tests are shown in Table \ref{tab:null_stats}. Both the full frequency spectrum null tests and the PB24 specific frequency null tests pass the criteria. Additionally, we scale the summary null statistics by the aliasing factor described in the previous section to validate frequencies between $1/2\Delta T$ and $3/2\Delta T$ and confirm that the null test pass criteria are still met.

\bgroup
\def\arraystretch{1.1}
\begin{table}[b]
\caption{\label{tab:null_stats}
PTE values of representative null statistics used in pass/fail criteria. ``DET'' refers to the detector based splits and ``OBS'' refers to observation based.
}
\begin{ruledtabular}
\begin{tabular}{lrr}
& \multicolumn{2}{c}{\textrm{PTE (\%)}} \\
\cmidrule{2-3}
&\textrm{DET}&\textrm{OBS}\\
\textit{Full spectrum null statistics}\\
Average $\chi^{\mathrm{DC}}_{\mathrm{null}}$ overall & 78.5 & 11.5\\
Average $\chi^{\mathrm{AC}}_{\mathrm{null}}$ overall & 12.1 & 2.9\\
Extreme $\chi^{2}_{\mathrm{null}}$ overall & 34.2 & 19.0\\
Extreme $\chi^{2}_{\mathrm{null}}$ by split & 24.2 & 11.5\\
Extreme $\chi^{2}_{\mathrm{null}}$ by frequency & 14.1 & 13.7\\
Total $\chi^{2}_{\mathrm{null}}$ & 8.5 & 10.6\\
\\
Lowest PTE & 33.8 & 11.8\\
Highest PTE & 64.7 & 99.4\\
\\
\textit{PB24 frequencies null statistics}\\
Average $\chi^{\mathrm{AC}}_{\mathrm{null}}$ overall & 3.6 & 74.2\\
Extreme $\chi^{2}_{\mathrm{null}}$ overall & 15.4 & 56.5\\
Total $\chi^{2}_{\mathrm{null}}$ & 6.7 & 33.8\\
\\
Lowest PTE & 9.3 & 58.4\\
Highest PTE & 98.1 & 48.3
\end{tabular}
\end{ruledtabular}
\end{table}
\egroup{}

\subsection{Systematic Correlation}
We test for possible systematic contamination by correlating the blinded Tau A polarization angles with various instrumental or environmental conditions. We then compare the correlation coefficients to correlation coefficients simulated using noise realizations to test for significance. We test a total of 11 potential systematic templates. It was using this method that the correlation with elevation/azimuth was discovered and motivated the polarization angle calibration described in Sec. \ref{sec:pol_angle_cal}. The correlations with all systematic templates investigated and their associated PTEs can be found in Table \ref{tab:syst_corr_pte}. We do not find any significant correlation (defined as PTE < 5\%) with potential systematics.

\bgroup
\def\arraystretch{1.1}
\begin{table}[b]
\caption{\label{tab:syst_corr_pte}%
PTEs of systematic template correlations
}
\begin{ruledtabular}
\begin{tabular}{lcc}
\textrm{Systematic template}&\textrm{Correlation Coefficient}&\textrm{PTE (\%)}\\
\colrule
PWV & -0.18 & 12.2\\
PWV/sin(elevation) & -0.15 & 17.1\\
Elevation & -0.01 & 93.1\\
Azimuth & 0.02 & 88.1\\
Outside temperature & -0.2 & 9.9\\
2f phase difference & -0.14 & 20.6\\
4f phase difference & -0.09 & 42.0\\
I2P amplitude & -0.13 & 26.5\\
I2P phase & -0.15 & 18.9\\
HWP vacuum & 0.03 & 78.1\\
Polarization fraction of Tau A & 0.07 & 53.0\\
\end{tabular}
\end{ruledtabular}
\end{table}
\egroup{}

\subsection{Systematic Errors}
We follow many of the detailed systematic studies presented in \citetalias{PB24}. Here we will note any differences between the analyses and will summarize the results for the studies kept identical. In addition to the studies presented in the \citetalias{PB24} analysis, we include a study for systematic polarization angle fluctuation due to our gain calibration. In Table \ref{tab:stat_and_syst_error}, we show the RMS of the estimated systematic errors as well as the median statistical errors per observation.

\subsubsection{Gain}
A time-varying misestimation of the relative gain between detectors may propagate into an error on the polarization angle measurement of Tau A. In order to estimate this effect, we utilize a second calibration pipeline where we use each detector's individual map of Tau A to calibrate its gain. We do this by fitting the Tau A signal in a detector's intensity map to a two-dimensional Gaussian model and using the amplitude of this fit to calibrate the detector's gain. We then estimate the systematic polarization angle fluctuation due to gain miscalibration as the difference in polarization angles between our nominal pipeline and this pipeline using Tau A for gain calibration. The systematic polarization angle fluctuation due to gain miscalibration is found to be 0.047$^{\circ}$ RMS.

\subsubsection{Intensity to polarization leakage}\label{I2P_systematic}
As noted in Sec. \ref{tod_processing}, the I2P leakage in this analysis is estimated on an observation-by-observation basis. In order to estimate the systematic error due to I2P leakage misestimation, we take the difference of Tau A polarization angles from pipelines using variable I2P leakage estimation and fixed I2P leakage estimates. We only include observations that pass our null tests using both I2P leakage models. The RMS of the angle difference between the two pipelines is $0.09^{\circ}$.

\subsubsection{Pointing}\label{pointing_systematic}
Misestimation of the boresight pointing will result in incorrect conversions between the polarization angle in telescope coordinates and sky coordinates. We evaluate this uncertainty by considering the observation-by-observation variation in the celestial position of Tau A as observed by PB-2a. Because we calibrate the pointing for individual observations using the intensity signal of Tau A, we consider this to be representative of random statistical fluctuations in our calibration. We find the median polarization uncertainty due to pointing fluctuation to be $0.002^{\circ}$.

\subsubsection{Residual $1/f$}
Correlated $1/f$ noise will be partially canceled during the bootstrap statistical uncertainty estimation process, and thus, isn't fully represented in the statistical uncertainty estimates. We estimate the residual $1/f$ uncertainty as follows. For both bootstrap noise maps and Tau A maps, we estimate $\delta Q =\sum_p^\mathrm{noise\,region}Q_p$ and $\delta U =\sum_p^\mathrm{noise\,region}U_p$ for 96 independent noise regions with the same size as the one used to estimate $d_t$ but outside of the Tau A region. We estimate the uncertainty in the polarization angle, $\sigma_{t}$, in the bootstrap maps and the Tau A maps using these noise samples. Assuming the noise regions of Tau A maps contain contributions from both statistical and residual $1/f$ noise whereas those of bootstrap maps contain contributions approximately from only statistical noise, we estimate the residual $1/f$ contribution by subtracting their variances following $\sigma_{1/f}=\sqrt{\sigma_\mathrm{Tau\,A}^2-\sigma_\mathrm{bootstrap}^2}$. The polarization angle fluctuation is calculated to be 0.016$^{\circ}$ RMS.

\subsubsection{Filtering and mapmaking bias}
The filtering and mapmaking pipeline will bias estimates of the Tau A polarization angle. A variation in these biases over time will produce a systematic polarization angle fluctuation. We estimate this systematic fluctuation with simulations. We simulate the Tau A map by convolving the IRAM Tau A map with a Gaussian beam with a FWHM identical to PB-2a's beam at 90 GHz. This serves as an input map as we simulate each detector's TOD for each observation by scanning the map, modulating the signal using HWP angles, and applying the same analysis method used on PB-2a data to estimate its angle. The systematic polarization angle fluctuation is calculated to be 0.026$^{\circ}$ RMS.

\bgroup{}
\def\arraystretch{1.1}
\begin{table}[b]
\caption{\label{tab:stat_and_syst_error}
Tau A polarization angle error budget. The median statistical uncertainty per observation and the RMS of the systematic polarization angle fluctuations are shown.
}
\begin{ruledtabular}
\begin{tabular}{lr}
\textrm{Type of statistical error}&\textrm{Median $\sigma_{t}$ ($^\circ$)}\\
\colrule
Statistical error of Tau A & 0.25 \\
Polarization angle calibration using $A_4$ & 0.02\\
Pointing calibration using Tau A intensity & 0.002\\
\toprule
\midrule
\textrm{Type of systematic error} & RMS ($^\circ$)\\
\colrule
Gain & 0.05\\
Ground & 0.01\\
I2P leakage & 0.09\\
Residual $1/f$ noise & 0.02\\
Time-domain filtering and mapmaking bias & 0.03\\
Detector polarization angle miscalibration & 0.03\\
\end{tabular}
\end{ruledtabular}
\end{table}
\egroup{}

\begin{figure*}
\includegraphics[width=\textwidth]{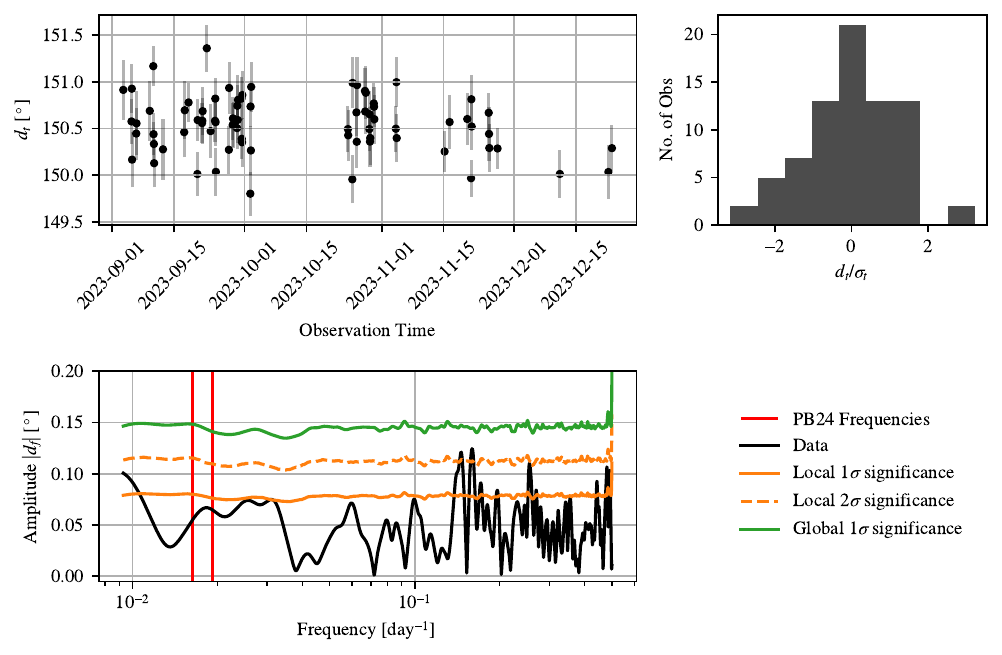}
\caption{\label{angle_tod}Top left: the measured polarization angles of Tau A during the 2023 observing season. Top right: The histogram of the measured polarization angles scaled by their assumed uncertainties. Bottom: Amplitude of the LSSA for frequencies less than 0.5 days$^{-1}$ as well as the approximate amplitudes of various levels of local and global significance. The two PB24 frequencies are called out with vertical lines.}
\end{figure*}

\begin{figure}
\includegraphics[width=\columnwidth]{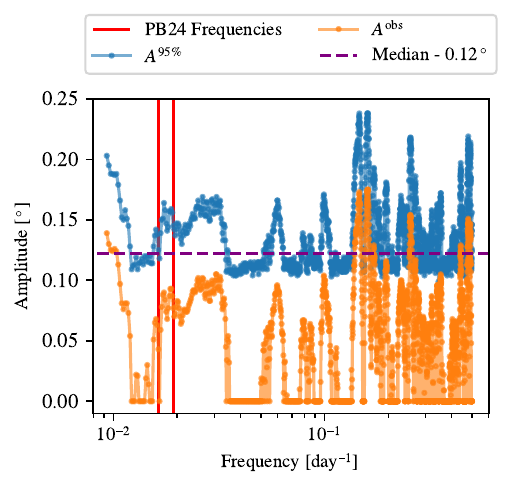}
\caption{\label{amp_spectrum}Observed polarization oscillation amplitudes and 95\% upper limits. The MLEs for the amplitude at the two PB24 frequencies are 0.04$^\circ$ and 0.08$^\circ$, respectively. The $95\%$ upper limits at the two PB24 frequencies are 0.13$^\circ$ and 0.15$^\circ$, respectively.}
\end{figure}

\subsubsection{Ground contamination}
We follow the method outlined in \citetalias{PB24} to assess the systematic error due to ground contamination in our maps of Tau A. We generate polarization maps of the ground signal by masking out Tau A in each observation. Concurrently, we generate hit maps of the pixels that Tau A passes through in ground coordinates. We use these hit maps and the season co-added ground polarization maps to estimate the per observation ground signal contamination into the estimated $Q$ and $U$ polarization of Tau A. We find that the systematic angle fluctuation due to ground signal contamination is 0.01$^\circ$ RMS.

\subsubsection{Detector polarization angle miscalibration}
A slightly different set of detectors is used for each observation in this analysis. Therefore, there may be a systematic fluctuation in the estimated polarization angle of Tau A due to averaging over detectors with varying amounts of polarization angle miscalibration. We estimate this systematic by estimating an effective miscalibration for each detector and then averaging these miscalibrations for the set of detectors used in the analysis of each observation. We find the fluctuation of the Tau A polarization angle due to detector polarization angle miscalibration to be 0.03$^\circ$ RMS.

\section{\label{sec:results}Results}

\begin{figure}
\includegraphics[width=\columnwidth]{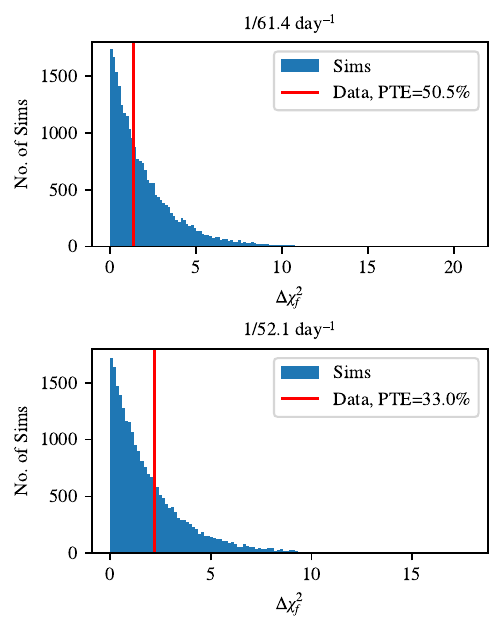}
\caption{\label{chi2_pb1_freqs}Distributions of the test statistic $\Delta\chi_f^{2}$ for noise only simulations at the two PB24 frequencies. The red line shows the local significance for $\Delta\chi_{f}^{2}$ in our data.}
\end{figure}

\begin{figure}
\includegraphics[width=\columnwidth]{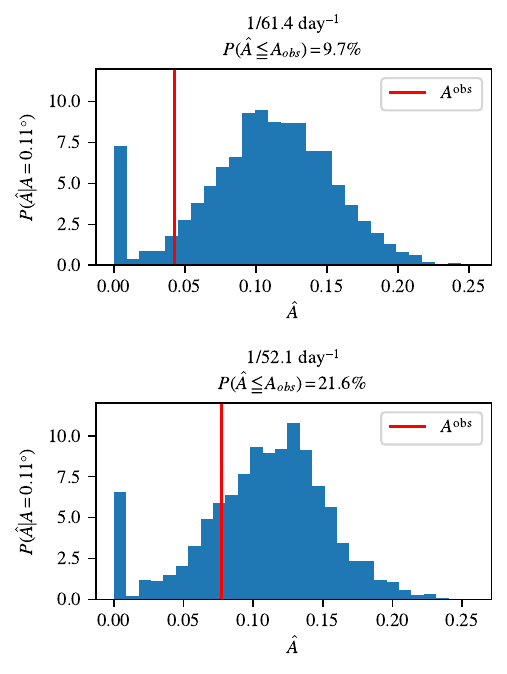}
\caption{\label{amp_MLE_likelihood}The probabilities\\
$P\left( \hat{A}\leqq A^{\mathrm{obs}} | A=0.11^\circ,f=1/61.4\ \mathrm{day}^{-1}\right)$ and\\
$P\left( \hat{A}\leqq A^{\mathrm{obs}} | A=0.11^\circ,f=1/52.1\ \mathrm{day}^{-1}\right)$ \\
that the MLE of A would be smaller than what was observed given the amplitudes of the oscillations reported in \citetalias{PB24}.}
\end{figure}

\begin{figure*}
\includegraphics[width=\textwidth]{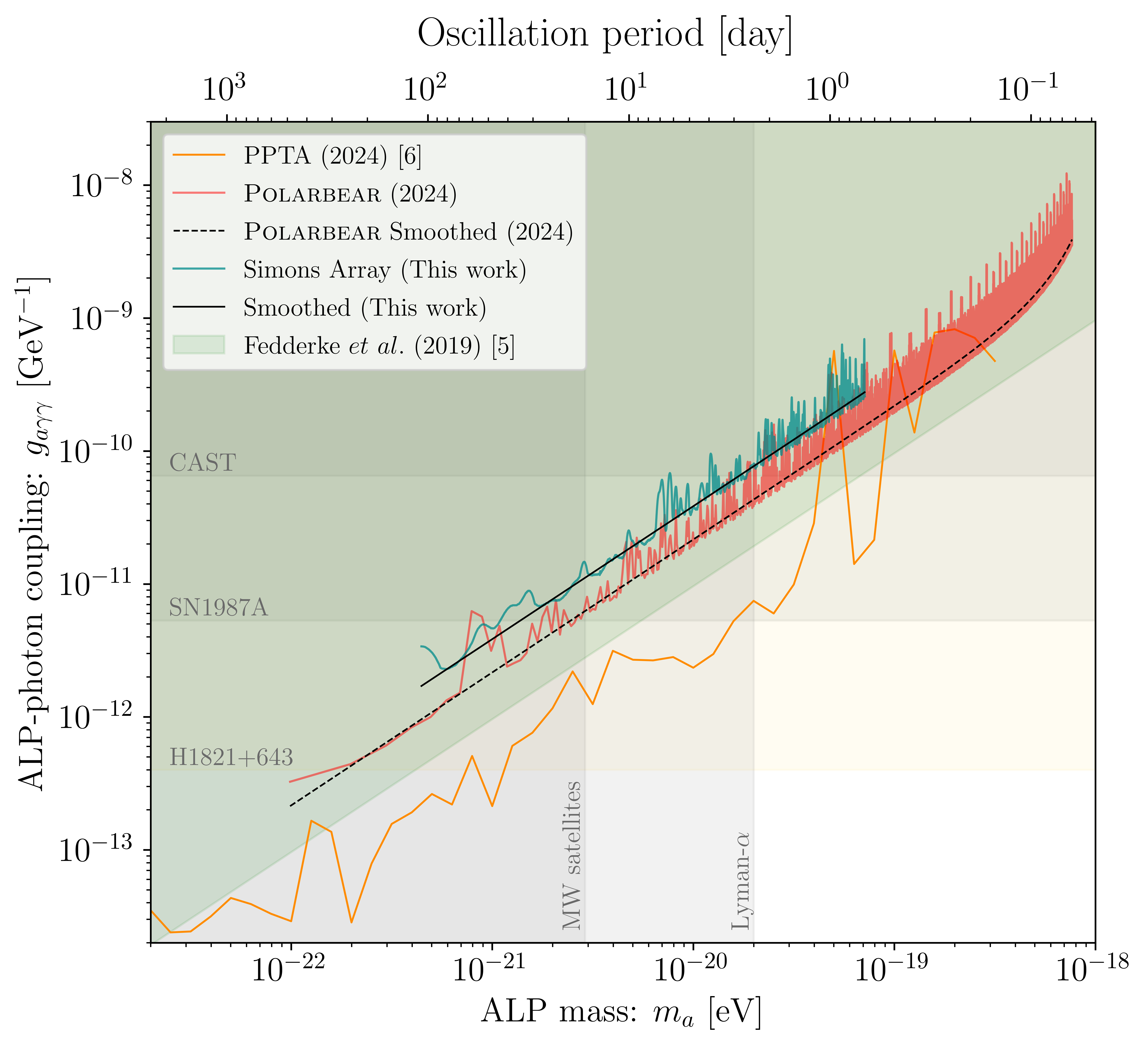}
\caption{\label{fig:coupling_constraint}95\% upper bound of $g_{a\gamma\gamma}$. The green curve represents the upper bounds of this study. The black curve represents the smoothed approximation of our bounds [Eq.~\ref{eq:g_a_bound_smooth}]. The red curve and black dashed curve represent the stochastic bound and its smoothed approximation set by \citetalias{PB24}. The orange curve represents the constraint from searching for axion-like oscillations from millisecond pulsars observations of the Parkes Pulsar Timing Array~\cite{PPTA}. The green shaded bound is from the lack of suppression of cosmic microwave background polarization due to axion-like oscillation in the recombination era~\cite{fedderke}. We show upper bounds from the CAST experiment~\cite{cast}, absence of excess of gamma-rays of SN1987A~\cite{sn1987a}, and absence of X-ray spectral distortions of quasar H1821+643~\cite{quasarh1821}. Additionally, we show lower bounds on the ALP mass by the Lyman-$\alpha$ forest~\cite{lyman_alpha_2} and the Milky Way satellite galaxies~\cite{mw_satellite_gal}.}
\end{figure*}

The timestream of the estimated Tau A polarization angles and its spectrum calculated via LSSA are shown in Fig.~\ref{angle_tod}. We simulate 24,000 noise realizations using the bootstrap method and show the estimated significance thresholds of the polarization angle oscillation, both locally and globally. We find no globally significant detection ($>3\sigma$) at any frequency. We calculate the observed maximum likelihood estimates (MLEs) for the polarization oscillation amplitudes as well as the $95\%$ upper limits using the Neyman procedure outlined in \citetalias{PB24}. The observed polarization angle oscillation amplitude, $A^{\mathrm{obs}}$, and the upper bound, $A^{95\%}$, can be seen in Fig.~\ref{amp_spectrum}. The median 95\% upper bound across the entire frequency spectrum is $0.12^\circ$. 

Additionally, we compute $\chi^{2}=\sum_t d^{2}_t/\sigma_t^{2}$ in the time domain to verify consistency. Here, we include the contribution of both statistical and systematic uncertainties. We find that this statistic yields a PTE of 0.14 compared to 24,000 noise-only realizations. Thus, the data are also consistent with the absence of a signal in the time domain.

We check for consistency between our results and those presented in \citetalias{PB24}. We note that there is no locally significant detection ($>3\sigma$) of polarization angle oscillation at either of the two PB24 frequencies in this analysis. We show the distribution of the simulated local test statistics at the two PB24 frequencies in Fig.~\ref{chi2_pb1_freqs} and highlight the test statistics calculated from our data. In order to perform a consistency check, we simulate 2000 realizations of noise and inject signals with amplitude $A=0.11^\circ$ at the PB24 frequencies. We note that there is some uncertainty in the frequency of the PB24 signals due to the finite duration of their dataset. This uncertainty is approximately 1/500 day$^{-1}$. The difference in time between the \citetalias{PB24} dataset and the observations in this analysis is $\approx$ 3000 days. The difference in the number of oscillations over the course of 3000 days for two frequencies separated by 1/500 day$^{-1}$ is $\approx$ 6 periods. Therefore, the uncertainty in the frequency means that we have no information about the phases of the potential signals during our observing period. Thus, we draw the phases of the simulated signals from a uniform distribution between 0 and 2$\pi$. We use these simulations to estimate $P \left(\hat{A} |A=0.11^\circ,f=1/61.4\ \mathrm{day}^{-1}\right)$ and $P \left(\hat{A} |A=0.11^\circ,f=1/52.1\ \mathrm{day}^{-1}\right)$. This distribution captures the probability density of the MLE of the oscillation amplitude, $\hat{A}$, given our estimated noise and a true signal with amplitude and frequency equal to that found in \citetalias{PB24}. These distributions for both PB24 frequencies, along with the observed MLEs for the oscillation amplitude, are shown in Fig.~\ref{amp_MLE_likelihood}. Using these distributions, we calculate $P\left( \hat{A}\leqq A^{\mathrm{obs}} | A=0.11^\circ,f=1/61.4\ \mathrm{day}^{-1}\right)$ and $P\left( \hat{A}\leqq A^{\mathrm{obs}} | A=0.11^\circ,f=1/52.1\ \mathrm{day}^{-1}\right)$. Their respective probabilities are $9.7\%$ and $21.6\%$. Thus, we cannot exclude the signals seen in \citetalias{PB24} at the 95\% level.

Additionally, we constrain the ALP-photon coupling. The $95\%$ upper bound of $g_{a\gamma\gamma}$ assuming the stochastic ALP field is also constructed using the Neyman procedure following the same method as \citetalias{PB24}, first calculating the constraint on the square of the effective amplitude of polarization oscillation, $g^2\equiv g_{a\gamma\gamma}^2 F^2_f\phi_\mathrm{DM}^2/4$ and then converting it to a constraint on $g_{a\gamma\gamma}$. Assuming that the ALP constitutes all of the local dark matter with a local dark matter density of 0.3 $\mathrm{GeV}/\mathrm{cm}^3$ and taking the median $95\%$ upper bounds on the effective oscillation amplitude below $f=1/2\Delta T$ of $\sqrt{g^{2,95\%}}=0.24^{\circ}$, the smoothed approximation of the bound is
\begin{equation}
    \begin{aligned}
    g_{a\gamma\gamma}\leq\left(3.84\times 10^{-12}\right)\times\left(\frac{m_a}{10^{-21}\,\mathrm{eV}}\right)\\
    \times\left[\mathrm{sinc}\left(\frac{m_a}{2.7\times10^{-19}\, \mathrm{eV}}\right)\right]^{-1}.
    \end{aligned}
\label{eq:g_a_bound_smooth}
\end{equation}
Note that the sinc function encapsulates the effect of averaging the Tau A polarization angle over a single observation in the transfer function $F_f$, as shown in Appendix A of \citetalias{PB24}. Our bounds are shown in Fig.~\ref{fig:coupling_constraint} along with other relevant constraints.

\section{\label{sec:conclusion}Conclusion}
We have used the 2023 season of Tau A observations from the PB-2a receiver to investigate the polarization variability of Tau A. We place a median 95\% upper bound on polarization angle oscillation at $A < 0.12^\circ$ for frequencies ranging from 3.39 year$^{-1}$ to 1.50 day$^{-1}$. To the best of our knowledge, this is the tightest bound placed on the polarization variability of Tau A at 90 GHz. Additionally, we investigate the hint of signal reported in \citetalias{PB24}. We detect no signal at either of the frequencies presented in the \citetalias{PB24} analysis. However, we do not exclude the signals seen in \citetalias{PB24} at the 95\% confidence level. When converting the polarization oscillation constraints to ALP constraints, we find a median 95\% upper bound on the ALP-photon coupling $g_{a\gamma\gamma}<\left(3.84\times 10^{-12}\right)\times\left(m_a/10^{-21}\,\mathrm{eV}\right)$ in the mass range from $4.4\times10^{-22}$ to $7.2\times10^{-20}$ eV.

The PB-2a telescope continued regular observations of Tau A through August 2024. We expect these observations to increase our data volume by a factor of $\approx$ 2 if the data selection efficiency remains consistent with that presented in this analysis. This would result in a sensitivity increase of $\approx$ 45\%. With this sensitivity, we expect to test the \citetalias{PB24} hint at the 95\% level. In addition to observing at 90 GHz, PB-2a observes in a band centered at 150 GHz. Further analysis including this second observing band will provide a check for frequency-dependent systematics, as any ALP-induced oscillation will be frequency-independent.

\begin{acknowledgements}
This work is supported by the Gordon and Betty Moore Foundation Grant No. GBMF7939, the Simons Foundation Grant No. 034079, the John Templeton Foundation Grant No. 58724 and the National Science Foundation grants AST-1440338 and AST-1207892. In Japan, this work is supported by the World Premier International Research Center Initiative (WPI Initiative) of MEXT, Japan, as well as JSPS KAKENHI grant Nos. 18H05539, 19H00674, and 23H00105, and the JSPS Core-to-Core Program JPJSCCA20200003. Work at LBNL is supported in part by the U.S. Department of Energy, Office of Science, Office of High Energy Physics under contract No. DE-AC02-05CH11231. Calculations were performed on the Central Computing System, owned and operated by the Computing Research Center at KEK. Carlo Baccigalupi acknowledges partial support by the Italian Space Agency LiteBIRD Project (ASI Grants No. 2020-9-HH.0 and 2016-24-H.1-2018), as well as the InDark and LiteBIRD Initiative of the National Institute for Nuclear Physics, and the RadioForegroundsPlus Project HORIZON-CL4-2023-SPACE-01, GA 101135036, and Project SPACE-IT-UP  by the Italian Space Agency and Ministry of University and Research, Contract Number  2024-5-E.0. Christian L. Reichardt acknowledges support from the Australian Research Council’s Discovery Project scheme (No. DP210102386). Masashi Hazumi acknowledges support from the National Science and Technology Council and the Ministry of Education of Taiwan. The Simons Array operated at the James Ax Observatory in the Parque Astronomico Atacama in Northern Chile under the stewardship of the Agencia Nacional de Investigaci\'on y Desarrollo (ANID). This research has made use of the data provided by the IRAM 30 meter telescope.
\end{acknowledgements}

\appendix
\section{Null test splits}\label{apx:null_test_splits}

This section describes the null test splits introduced in Sec.~\ref{sec:data_validation}. For the following detector splits, ``A/B pixel handedness'', ``Q/U type pixel'', ``top/bottom detector'', ``left/right on focal plane'', and ``upper/lower on focal plane'', the data is split according to a discrete property of the detector, and thus, there won't necessarily be exactly equal data volume in each half of the split. However, for the remaining detector and observation based splits, the data is separated at the median value of the split parameter.

The detector based splits are
\begin{itemize}
\item ``A/B pixel handedness": the sinuous antennas on the focal plane contain two types of pixel handedness, referred to as A and B, that are mirror images of each other. We split the detectors based on their associated antennas' handedness to check for systematics related to the polarization angle frequency dependence of the sinuous antenna.
\item ``Q/U type pixel": there are two sets of sinuous antenna polarization angles on the focal plane. We split the data between the types to check for problems in device fabrication.
\item ``top/bottom detector": there is a pair of detectors for each band associated with a single antenna. Each detector is sensitive to both orthogonal polarization states when utilizing the rotating HWP. We split the data along this property to test for device mismatch.
\item ``left/right on focal plane" and ``upper/lower on focal plane": we split the data based on the physical location of the detector on the focal plane to check for optical systematics or systematics confined to specific detector wafers.
\item ``gain": we split the detectors based on the gain calibration factor to probe for problems associated with the calibration.
\item ``I2P amp": we split the data based on the estimated amplitude of the I2P leakage to probe for systematic contamination coupling through I2P leakage.
\item ``4f amp/2f amp": we split the data based on the amplitude of the 2nd and 4th harmonics of the HWP synchronous signal. This probes for systematics related to the amplitude of instrumental polarization or optical systematics observed by different detectors.
\item ``4f phase/2f phase": we split the data based on the phase of the 2nd and 4th harmonics of the HWP synchronous signal. This probes for issues related to the per-observation polarization angle calibration.
\item ``white noise": we split the data based on the white noise level of the detectors to probe for ineffective data selection or noise modeling.
\end{itemize}

The observation splits are

\begin{itemize}
    \item ``1st/2nd half of season": we split the observations chronologically into halves to probe for time-dependent miscalibrations or changes in the instrument.
    \item ``PWV": we split the observations based on the PWV during the observation to probe loading or weather-dependent systematics.
    \item ``sun distance/moon distance": we split the observations based on the angular distance to the sun/moon. This probes for potential optical systematics caused by far-sidelobe contamination.
    \item ``elevation": we split the observations into high and low elevation observations to check for residual systematics associated with the observing orientation.
    \item ``4f phase/2f phase": we split the observations based on the median HWP synchronous signal 2f/4f phases over all active detectors in the observation. This is sensitive to observations with residual systematic polarization angle miscalibrations.
    \item ``4f amp": we split the observations based on the median amplitude of the HWP synchronous signal 4f amplitude over all active detectors in the observation. This probes for systematic differences in observations with varying instrumental polarization.
    \item ``white noise": we split the observations based on the median white noise level over all active detectors in the observation. This is sensitive to observations with anomalous noise that may contaminate the polarization angle estimate.
    \item ``gain": we split the observations based on the median gain over all active detectors in the observation. This is sensitive to systematic gain miscalibrations.
    \item ``I2P amplitude": we split the observations based on the median I2P amplitude over all active detectors in the observation. This probes for systematic I2P leakage misestimation.\\
    \item ``HWP vacuum": The HWP vacuum split refers to the pressure in the vacuum system used to adhere the anti-reflection coating on the birefringent sapphire optical surfaces. This vacuum level naturally degrades over time and is regularly pumped out during maintenance. We split the observations based on the vacuum level of the HWP system during the observation. This probes for potential systematics associated with HWP optical stack.
\end{itemize}

\nocite{*}

\bibliography{citations}

\end{document}